\documentclass{aip-cp}

\usepackage[numbers]{natbib}
\usepackage{rotating}
\usepackage{graphicx}

\begin{document}

\title{Constant temperature description of the nuclear level densities}

\author{Mihai Horoi\corref{cor1}}
\author{Jayani Dissanayake}
\eaddress{dissa1wa@cmich.edu}

\affil{Department of Physics,
Central Michigan University,
Mount pleasant, MI 48859, USA}
\corresp[cor1]{Corresponding author: mihai.horoi@cmich.edu}

\maketitle

\begin{abstract}
The spin and parity dependent nuclear level densities (NLD) are calculated for medium-heavy nuclei using shell model techniques. The NLD are used to calculate cross sections and reaction rates of interest for nuclear astrophysics and nuclear energy applications. We investigate a new approach of describing the shell model NLD via a constant temperature parametrization. This approach provides new information about the effects of symmetries on the temperature of the low-lying nuclear states, and it is shown to be more versatile for applications.
\end{abstract}

\section{INTRODUCTION}

The nuclear level densities are important quantities that provide interesting information about the nuclear structure properties of the atomic nuclei \cite{95ho519,95ze141,96ze85}. They are very useful ingredients for the calculation of the compound nucleus cross sections \cite{52ha366,07vo044602} of interest for nuclear astrophysics and nuclear engineering applications. In addition, the level densities are necessary for the partition functions entering the calculation of the reaction rates for nuclei in hot stellar environments \cite{10ho222}.

Over the years we have developed a methodology \cite{13se215,11se413,10sc520,10se024,07ho262,06ho107,05ho142,04ho041,03ho054} for calculating the shell model spin- and parity-dependent nuclear level densities using methods of statistical spectroscopy \cite{71fr94,86wo1}.   A number of important applications were also reported in Proceedings of different conferences \cite{06ho120,08ho132,10ho222,16ze012,14se}. Our method is based on the first two moments, the centroids and the widths, of the nuclear configurations contributing to the low-energy distributions. The results compare very well with those of full shell model diagonalization of the same shell model Hamiltonian, which in some cases can be calculated using shell model codes, such as NuShellX \cite{nushellx}. We found out that the shell model NLD calculated with nuclear Hamiltonians fine-tuned to describe the low-energy nuclear spectroscopy, such as USDA \cite{usda} for the sd-shell nuclei or GXPF1A \cite{gxpf1,gxpf1a} for the fp-shell nuclei describe reasonably well the available level counting data for some sd-shell nuclei, such as $^{26}$Al and $^{28}$Si, for which many low-lying levels are known. For the fp-shell nuclei one also gets good agreement with some of the low-energy neutron resonance data maintained by RIPL \cite{ripl}. 
Others approaches to the nuclear level densities based on nuclear shell model Hamiltonians can be found in Refs.\cite{05eg044311,05eg067304,09eg054310,12eg012028,16se064304,15al171,06te067302}. 

The nuclear level densities can be used to understand the atomic nuclei as mesoscopic systems by looking to their ergodic properties. In Refs. \cite{95ho519,95ze141,96ze85} we investigated these properties. For applications in nuclear astrophysics and nuclear engineering one needs to properly use the NLD as input to nuclear reaction codes, such as TALYS \cite{talys16}. In the past \cite{10ho222} we used our shell model NLD input as tables. This approach is difficult to implement in TALYS and it has other shortcoming related to the continuation in excitation energy of the NLD beyond 10-12 MeV. 
Given the recent success of the constant temperature parametrization of the experimental data on nuclear density of states \cite{05eg044311,09eg054310,12eg012028} we decided to investigate whether a constant temperature parametrization is appropriate for the shell model spin- and parity-dependent nuclear level densities. 

The paper is organized as follows. The next section gives a brief description of our moments method approach to calculate the spin- and parity-dependent shell model NLD, which is followed by a description of the constant temperature parametrization of the NLD. The final two sections are devoted to the results and conclusions.



\begin{figure}[h]
  \centerline{\includegraphics[width=350pt]{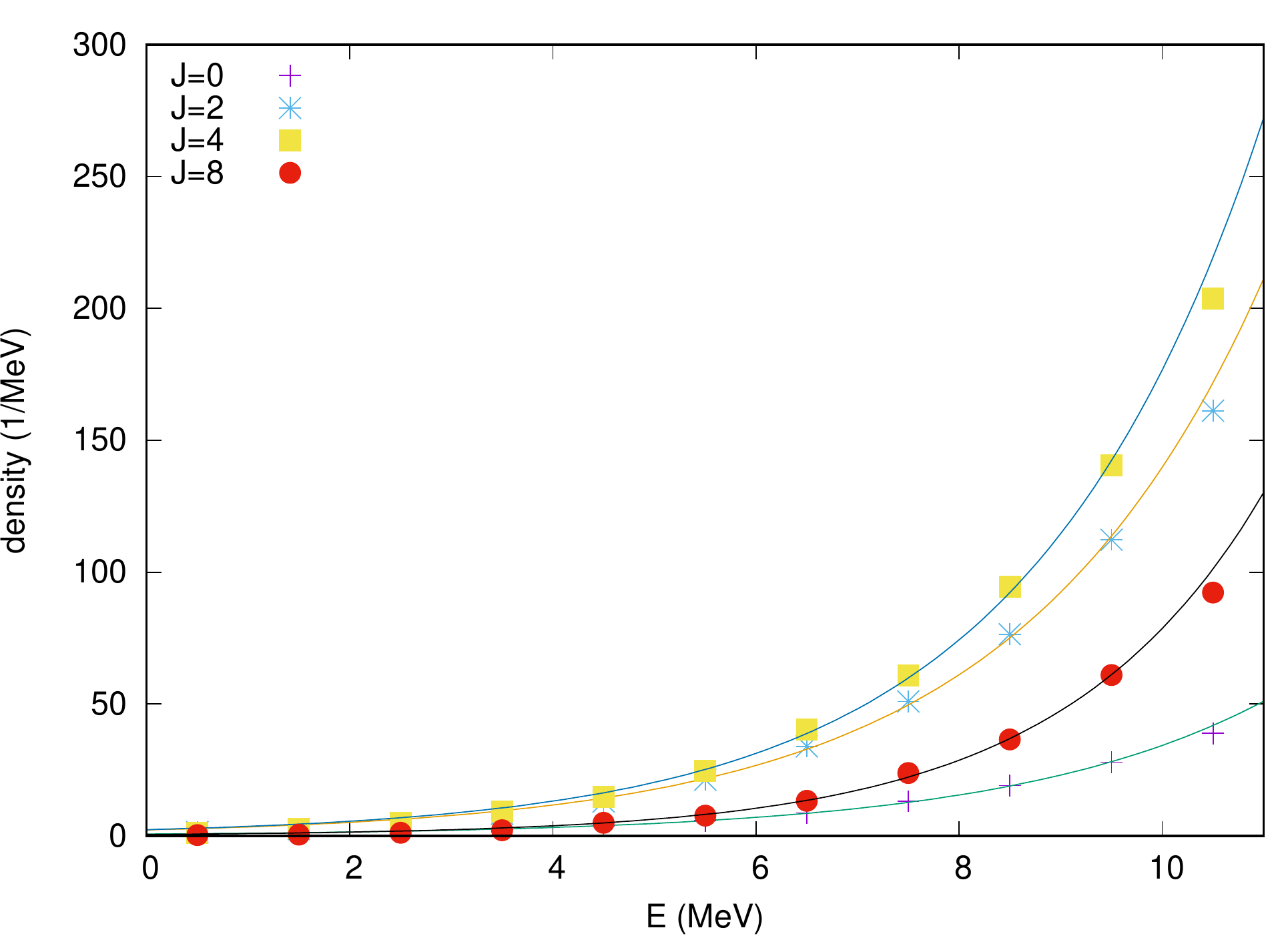}}
  \caption{$^{62}$Fe NLD vs energy for several $J$-values. The symbols are the results of the moments method, and the lines are the fitting by the constant temperature formula, Eq. (\ref{moroj}).}
\end{figure}

\section{MOMENTS METHOD NUCLEAR LEVEL DENSITY}

We start following closely the approach proposed in Refs. \cite{04ho041,11se413}. For reader`s convenience, we first repeat the main equations of the moments method. According to this approach, one can calculate the level density $\rho$ for a given spin $J$ and a given parity $\pi$ as a function of excitation energy $E_x$ in the following way:

\begin{equation}
\rho\left( E_x,J,\pi \right) = \sum_{c\ \in \ configs} D_c(J,\pi)\ G_{FR} \left( E_x, E_c(J),\sigma_c(J) \right) \ ,
\label{mdens}
\end{equation}
where $c$ are configurations of protons and neutrons in a set of valance spherical orbitals. Examples of such valence spaces are the $sd$ shell consisting of the $0d_{5/2}$, $0d_{3/2}$ and $1s_{1/2}$ orbitals, or the $fp$ shell consisting of the $0f_{7/2}$, $0f_{5/2}$, $1p_{3/2}$ and $1p_{1/2}$ orbitals. The $E_c(J)$ and $\sigma_c(J)$ are the fixed-J centroids and widths, respectively,  corresponding to the configurations $c$ of nucleons distributed in the corresponding valence space orbitals. They can be calculated without performing a full diagonalization in the valence space, by calculating traces of the first two powers of the Hamiltonian:

\begin{equation}
E_c(J),\ \sigma_c(J)\ \longleftarrow \ Tr_{SD_c} \left< M \mid H^q \mid M \right>_{SD_c}\ .
\label{mcewi}
\end{equation}
Detailed expression for these traces can be found in Refs. \cite{04ho041,11se413}. A high performance computer implementation of the algorithm used to calculate these moments can be found in \cite{13se215}. Here $E_x=E-E_{g.s.}$, where $E_{g.s.}$ is the ground state (g.s.) energy. This implies that a good estimation of the g.s. energy is required. This estimation is usually done by direct shell model diagonalization, or using extrapolation techniques, such as the exponential convergence method \cite{99ho206,02ho027,03ho034}, or the increased truncation method described in Ref. \cite{11se413}.



\begin{figure}[h]
  \centerline{\includegraphics[width=350pt]{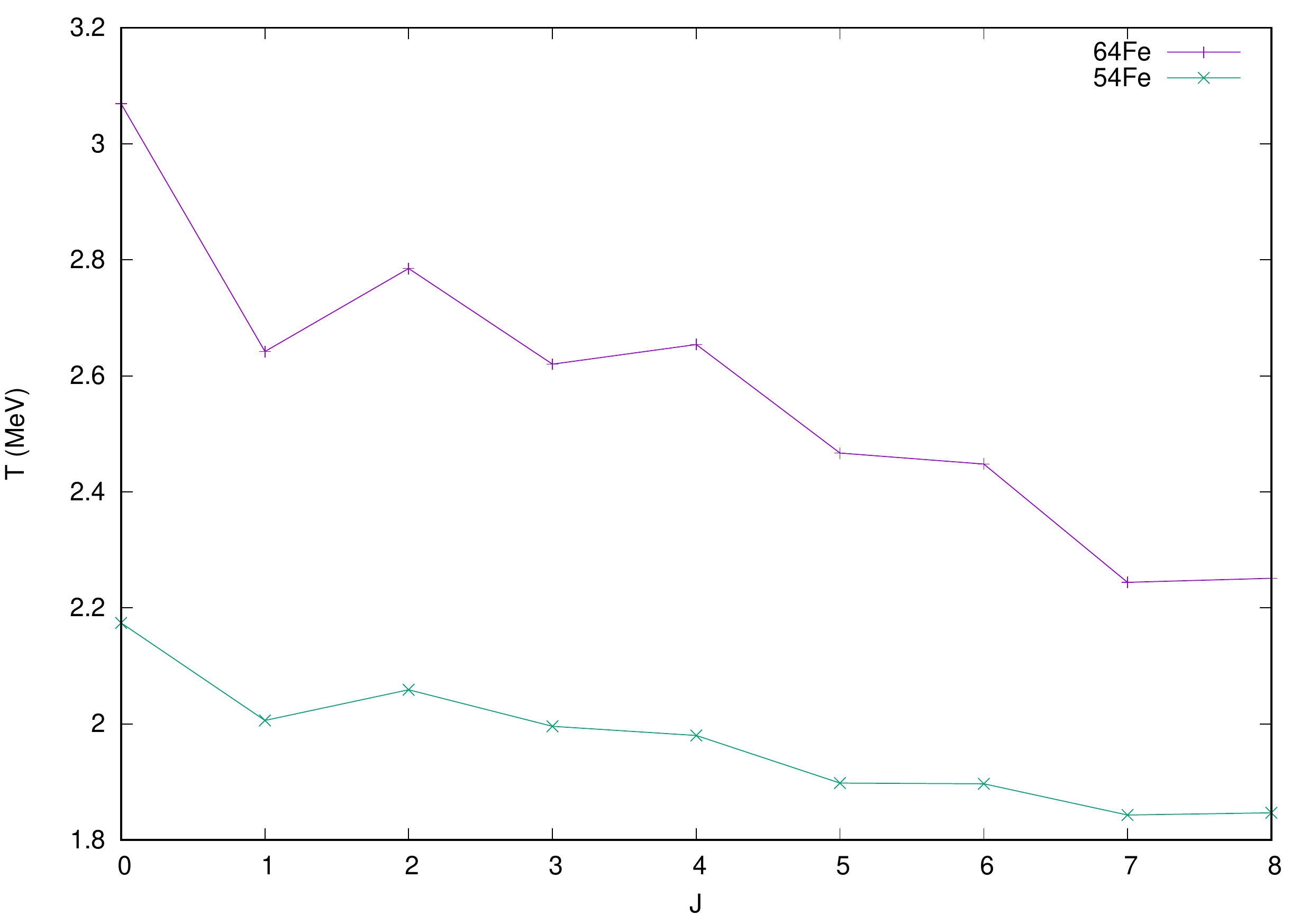}}
  \caption{Temperature vs $J$ for two iron isotopes.}
\end{figure}

\section{CONSTANT TEMPERATURE PARAMETRIZATION}

Given the large interest for accurate nuclear level densities, several parametrizations of the density of states were proposed over the years. Among them, the back-shifted Fermi gas formula \cite{65gi1446} and the constant temperature formula \cite{65gi1446,05eg044311}.
 In the von Egidy and Bucurescu \cite{09eg054310,12eg012028} approach the constant temperature density of states is given by 

\begin{equation}
\rho_{CT}(E_x)=\frac{1}{T}e^{(E_x-E_0)/T}
\label{eqros}
\end{equation}
Then, the J-dependent nuclear level density can be calculated by the approximate formula

\begin{equation}
\rho_{CT}(E_x,J)=\frac{2J+1}{2\sigma^2}e^{-J(J+1/2)/(2\sigma^2)}\rho_{CT}(E_x)\ .
\label{eqroj}
\end{equation}
where $\sigma$ is an appropriate spin cutoff parameter that depends on the excitation energy and other isotope specific factors.
Here, the assumption is that both parities have the same nuclear level density.
This approach has the disadvantage that constant temperature applies only to the density of states, and therefore the temperature is not constant for each J (or parity). 

The constant temperature approach success can be understood as due to a pairing first order phase transition. The semiclassical BCS description of the pairing is consistent with a second order phase transition. However, higher nonlinearities induced in many body effective Hamiltonians could lead to first order pairing phase transitions \cite{07ho054,87ap197,87ap64,90du653}. These many body interactions can be originating from full shell model Hamiltonian projected on restricted model spaces, such as those included in the BCS approach. A detailed connection between these points of view is not yet available, but seems quite plausible. Additional arguments in favor of an exponential (rather than a Gaussian) tail of the distributions described by shell model Hamiltonian can be extracted from Figs. 9-10 of Ref. \cite{96ze85}.  

Although in the shell model the phase transitions are not sharp, one can see that they are likely to depend on the symmetries left \cite{07ho054}. Therefore, here we consider a constant temperature formula for each angular momentum $J$ and each parity $\pi$:

\begin{equation}
\rho_{CT}(E_x,J,\pi)=\frac{1}{T_{J\pi}}e^{(E_x-E_{0J\pi})/T_{J\pi}}
\label{moroj}
\end{equation}
This parametrization of the spin- and parity-dependent NLD is similar to that of Eq. (\ref{eqros}), except that the temperatures $T_{J\pi}$ and the energy shifts $E_{0J\pi}$ are expected to be different for each parity and spin. It is expected to better describe an ergodic system without mixed symmetries.

\begin{figure}[h]
  \centerline{\includegraphics[width=350pt]{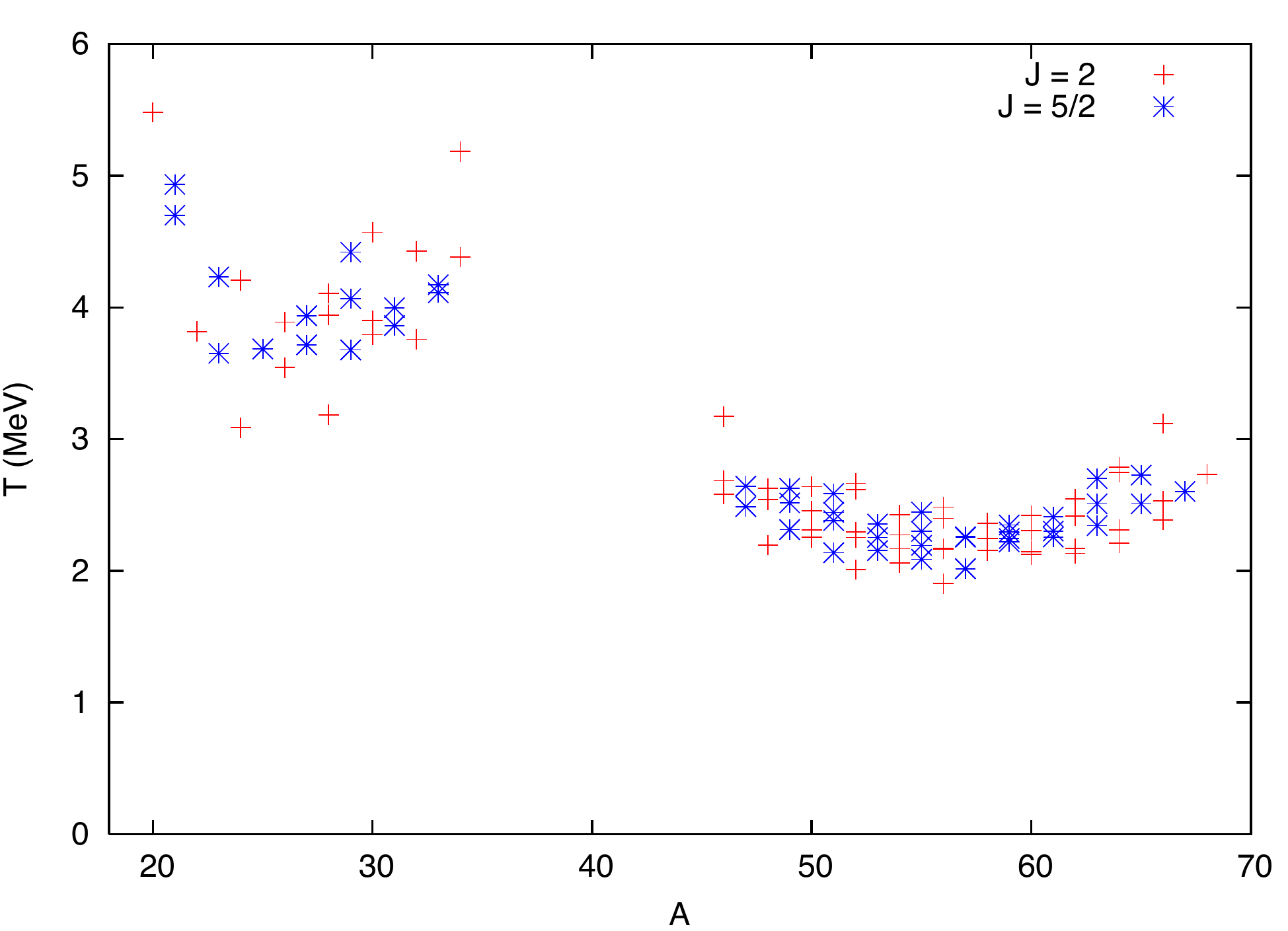}}
  \caption{Temperature vs the mass number A for some $sd$- (left) and $fp$-shell (right) nuclei.}
\end{figure}

\subsection{Results for sd- and fp-shell nuclei}

We calculated the moments method NLD for a large number of sd-shell and fp-shell nuclei using the USDA and the GXPF1A Hamiltonians, respectively. In addition, we fitted the moments NLD to the spin and parity constant temperature formula, Eq. (\ref{moroj}). The results are analyzed below.

Figure 1 shows the NLD of $^{62}$Fe calculated for $J=0,\ 2,\ 4$, and 8. The symbols are the results of the moments method, and the lines are the fitting by the constant temperature formula, Eq. (\ref{moroj}). The calculations were done in the $fp$ valence space using the GXPF1A Hamiltonian. The slope of different curves indicates different temperatures for different symmetries (total angular momenta), while Eqs. (\ref{eqros}) - (\ref{eqroj}) are using the same temperature for all $J$s.

Figure 2 shows the temperature for two isotopes as a function of the total angular momentum. This results clearly indicate that for the same nucleus the temperature is dependent on the total angular momentum. In addition, one can see that different isotopes of the same element having different masses could exhibit significantly different temperature characterizing the low-lying spectra. In addition, one can observe the odd-even $J$ staggering typical for the pairing effects in even-even nuclei (see e.g. Figs. (5)-(6) of Ref. \cite{07ho054}).

Figure 3 shows the spin and parity temperature as a function of the mass number A for series of even and odd mass sd-shell and fp-shell isotopes. For the even mass isotopes the selected total angular momentum is 2 (and positive parity), and for the odd mass isotopes the selected total angular momentum was 5/2 (the parity was positive for the sd-shell nuclei at smaller A, and negative for the fp-shell nuclei at larger A). The typical increase of the temperature towards the shell closure, at A=16 and A=40, was also observed in the data \cite{09eg054310,12eg012028}. The general decrease of the temperature with the mass number observed in \cite{09eg054310,12eg012028}, is also present in our results. The temperatures tend to stay in narrow bands for a large number of isotopes of different masses, indicating that a parametrization and extrapolation of the quantities to other nearby isotopes could be possible.

Figure 4 shows the spin and parity energy shift $E_{0J\pi}$ as a function of the mass number A for the same isotopes, angular momenta, and parities as in Fig. 3. The energy shifts tend to be much more scattered than the temperature in Fig. 3, suggesting that a better understanding of their larger range is needed before considering any extrapolations.

\begin{figure}[h]
  \centerline{\includegraphics[width=350pt]{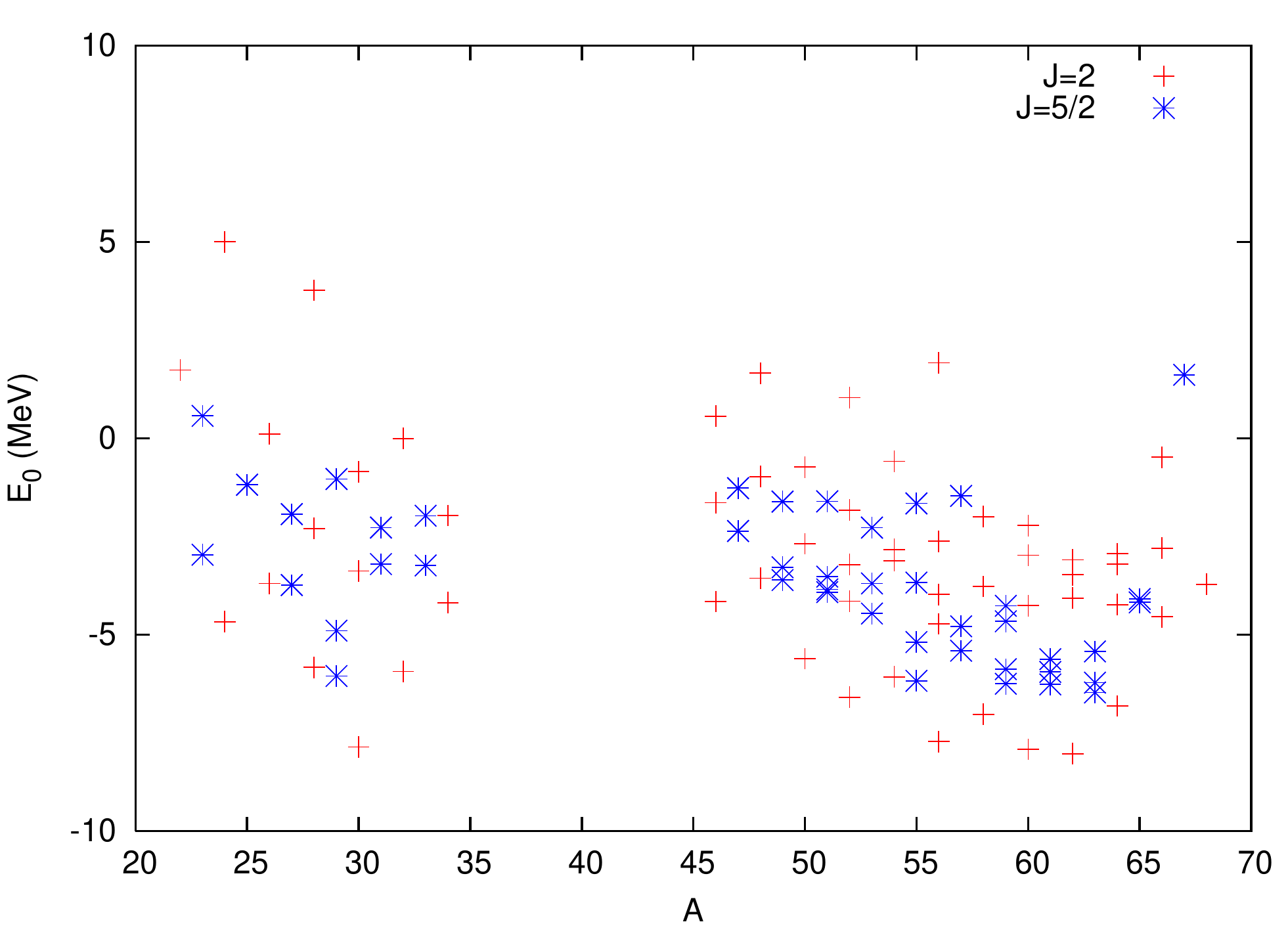}}
  \caption{Energy shifts $E_0$ vs the mass number A for some $sd$- (left) and $fp$-shell (right) nuclei.}
\end{figure}




\section{CONCLUSIONS AND OUTLOOK}

In conclusion, we studied the moments method fixed spin and parity nuclear level densities for a large number of $sd$- and $fp$-shell nuclei using realistic shell model Hamiltonians. We also investigated the adequacy of the constant temperature approximation for these shell model densities, and we argued that a constant temperature approach to these densities may be better justified than for the density of states, which includes subsets from different symmetry representations (e.g. different spins).

We found that indeed, the temperatures corresponding to different total angular momenta of the same isotope could be quite different. We also showed that these temperatures exhibit and odd-even spin effect similar to that found in the pairing strength of the low-lying states of nuclei. This feature supports the conjecture that the constant temperature description of the low excitation energy nuclear level density could be explained by a pairing phase transition at almost constant temperature. However, more investigations need to be done to fully establish this connection.

Finally, we investigated the behavior of the parameters entering the constant temperature formula Eq. (\ref{moroj}),  namely the temperature and the energy shift, and we found that one could potentially extrapolate these parameters to other neighboring isotopes of interest. This outcome could be very useful for a simple integration of the results of Eq. (\ref{moroj}) within reaction codes, such as TALYS.


\section{ACKNOWLEDGMENTS}
Support from the U.S. NSF Grant No. PHY-1404442 and the NUCLEI SciDAC Collaboration under U.S. Department of Energy Grant No. DE-SC0008529 is acknowledged.


\bibliographystyle{apsrev}%
\bibliography{nmh,misc,nld}%

\end{document}